\documentclass[12pt]{article}
\usepackage[dvips]{graphics}
\title{On particle production by classical backgrounds}
\author{Hrvoje Nikoli\'c \\
Theoretical Physics Division, Rudjer Bo\v{s}kovi\'{c} Institute, \\
P.O.B. 180, HR-10002 Zagreb, Croatia \\
{\normalsize hrvoje@faust.irb.hr} \\
\makebox[1in]{} \\
}
\date{\today}
\begin{document}
\maketitle
\begin{abstract}
The theory of quantum fields in classical backgrounds is re-examined 
for the cases in which the Lagrangian is quadratic in quantum fields. 
Various methods that describe particle production are discussed. 
It is found that all methods suffer from certain ambiguities, 
related to the choice of coordinates, gauge, or  
counter-terms. They also seem to be inconsistent with the 
conservation of energy. This suggests that such classical 
backgrounds may not cause particle production. 
\end{abstract}
\vspace*{0.5cm} 
\vspace*{0.9cm}


\noindent
One of the main differences between the first quantization and 
the second quantization (quantum field theory) is the fact 
that the latter is able to describe production and destruction 
of particles. A particularly interesting, but still 
experimentally unverified prediction of quantum 
field theory is that the vacuum may be unstable, i.e. that a 
nontrivial classical background may cause particle production. 
For example, such an effect is believed to exist in the    
electromagnetic \cite{schw} and gravitational \cite{bd}  
background.   
However, the concept of particle is not a 
fundamental concept in quantum field theory. It is well defined 
only for free fields, i.e. fields described by a free Lagrangian. 
On the other hand, production and destruction of particles may 
occur only if an interaction is present. Therefore, it is 
convenient to write the Lagrangian as ${\cal L}={\cal L}_{\rm free}
+{\cal L}_{\rm int}$ and use the interaction picture. 
In the general case, it is not clear how to 
separate the ``free" part from the ``interaction" part in a unique way.  
As we demostrate below, this is not merely a matter 
of convenience in some cases. 
Instead, the physical predictions related to 
particle production may depend on what we mean by ``free" 
particles. 

In this letter we study systems described by Lagrangians 
quadratic in a quantum field $\phi$. Such Lagrangians can be 
written in an elegant way as   
\begin{equation}\label{quadr}
{\cal L}=\phi^{\dagger}J\phi \; .
\end{equation}
Here $J$ can be a c-number function or a derivative operator, 
but not a quantum-field operator. To clarify the meaning of 
$J$, we give a few examples. If $\phi$ is a free scalar complex 
field, then $J= {\stackrel{\leftarrow\;}{\partial^{\mu}}}
\partial_{\mu} -m^2 $. (The arrow denotes that the operator 
acts on the left.)
The generalization of this 
for a curved background described by the metric $\bar{g}_{\mu\nu}$ is 
$J= {\stackrel{\leftarrow\;}{\partial^{\mu}}} 
|\bar{g}|^{1/2} \partial_{\mu} -m^2 $. If $\phi$ is a Dirac spinor 
interacting with the electromagnetic background $\bar{A}_{\mu}$, 
then $J=\gamma^0 [i\gamma^{\mu}(\partial_{\mu}+ie\bar{A}_{\mu})-m]$.
 
The quantity $J$ can always be written as 
\begin{equation}\label{Jfi}
J=J_{\rm free}+J_{\rm int} \; ,
\end{equation} 
which defines the interaction picture and the 
corresponding Feynman rules. The elementary vertex takes the form 
as in Fig. 1. There are three possible physical interpretations of 
the diagram in Fig. 1. The first interpretation is the renormalization 
of the free propagator. The second interpretation is scattering 
caused by the classical background $J_{\rm int}$. 
The third interpretation is pair production induced 
by the classical background $J_{\rm int}$. We argue below that 
the third interpretation may not be physical. Irrespective  
of the interpretation, 
the most general diagram can be represented as a combination of 
disconnected diagrams with two external legs. The diagram with 
two external legs calculated at all orders of perturbation 
theory is represented by Fig. 2. It is important to note that, 
for a given $J$, the separation (\ref{Jfi}) is not unique. 
Let us discuss a few examples. 

We begin with the case 
$J_{\rm free}= {\stackrel{\leftarrow\;}{\partial^{\mu}}}
\partial_{\mu}$, $J_{\rm int}=-m^2$. In this case, the most 
natural interpretation of Figs. 1 and 2 is the renormalization 
of the propagator. Indeed, the diagrams of Fig. 2 can be easily 
summed \cite{ryder}, leading to the usual massive propagator 
represented by the left-hand side of Fig. 2. 
The scattering interpretation is also meaningful. If we define scattering 
as a deviation of the particle trajectory from the trajectory 
that would be realized if the particle were ``free", then 
it becomes clear that the mass term can be interpreted in such a 
way, simply because the trajectory of a massive particle is different 
from that of a massless particle.            
Pair production is also possible, in the sense that the amplitude 
for the production of a particle-antiparticle pair does not 
vanish. However, real massless particles cannot be produced 
simply because it is forbidden {\em kinematically}. Such a production 
would not be consistent with the conservation of energy. 

Let us now discuss a more interesting case, in which 
$J_{\rm free}={\stackrel{\leftarrow\;}{\partial^{\mu}}} 
\partial_{\mu} -m^2 $ (for bosons) or 
$J_{\rm free}=\gamma^0 (i\gamma^{\mu}\partial_{\mu}-m)$ 
(for fermions). This corresponds to the usual definition 
of the ``free" Lagrangian. We assume that $J_{\rm int}$ is a 
non-trivial function, corresponding to a gravitational, 
electromagnetic, or any other background field. In this case, a   
natural interpretation of Figs. 1 and 2 is the scattering 
interpretation. The propagator-renormalization interpretation 
is also possible. Instead of summing the diagrams of Fig. 2, 
the renormalized propagator can be found more directly in the 
following way: (We present the method for bosons, while that 
for fermions is essentially the same.)
The Lagrangian (\ref{quadr}) provides that 
the equation of motion for $\phi$ is a linear 
homogeneous equation, so the solution can be expanded as        
\begin{equation}\label{expan}
\phi(x)=\sum_k a_k f_k(x) + b^{\dagger}_k g^*_k(x) \; .
\end{equation}
The operators $a_k$, $a^{\dagger}_k$, $b_k$ and $b^{\dagger}_k$ 
satisfy the usual algebra of raising and lowering operators. This
allows us to introduce the state $|0\rangle$ with the property
$a_k|0\rangle=b_k|0\rangle=0$. The propagator $G(x,y)$ is then 
given by 
\begin{equation}\label{prop}
G(x,y)=\langle 0|O \phi^{\dagger}(x)\phi(y)|0\rangle \; ,
\end{equation}
where $O$ is an ordering operator, depending on which propagator 
we want to obtain. For example, the Feynman propagator 
corresponds to the time ordering. 

The crucial question is whether the diagrams in Figs. 1 and 2 
may correspond to pair production in the case
of non-trivial $J_{\rm int}$. The 
corresponding amplitude does not vanish. However, the process 
is kinematically forbidden again, because the 
production of particles increases the energy of the system. 
Nevertheless, the usual argument that supports the 
pair production is a claim that the pair production causes 
a backreaction on the ``source" $J_{\rm int}$, such that 
the energy is conserved. To avoid a possible misunderstanding, 
note that this backreaction cannot correspond to a 
recoil of a massive object that generates $J_{\rm int}$, because 
the recoil only saves the conservation of the 3-momentum, not 
the conservation of energy. Therefore, the physical 
mechanism that causes this hypothetical backreaction is not 
clear. The backreaction is not a consequence of the basic 
principles of quantum field theory. If one insists on retaining the 
possibility of pair production, then the backreaction 
should be included in a vague way, by hand. 
If the backreaction really exists, then it is not clear why it does 
not exist for the case $J_{\rm int}=-m^2$ discussed above. 
For example, particles could be spontaneously created in  
flat space-time, which would be accompained by a spontaneous 
change of the background metric, such that the negative 
gravitational energy would cancel the positive energy of 
the particles created. Of course, the experimental observation  
that particles are not spontaneously created in flat 
space-time is not a surprise. Our point is that,  
for any $J_{\rm int}$, the 
existence of the backreaction does not seem to be reasonable. 
Therefore, the conclusion that $J_{\rm int}$ does not 
cause pair production seems to be the most reasonable. 

There are also other arguments against pair production, 
which do not rely on energy conservation. 
Let us pay particular attention to the case of electromagnetic 
background. In this case, the full Lagrangian ${\cal L}$ is 
gauge invariant, but its separate parts ${\cal L}_{\rm free}$ and 
${\cal L}_{\rm int}$ are not. Therefore, the physical results 
that follow from the calculation of the diagrams in 
Figs. 1 and 2 depend on the choice of gauge. Of course, 
the renormalized propagator (\ref{prop}) should depend on gauge. 
However, the number of produced particles should not depend 
on gauge. If the calculation is based on the 
diagrams in Figs. 1 and 2, then the only gauge-invariant 
conclusion related to pair production is that these diagrams 
do not describe pair production. 

The particle production by an electromagnetic background 
can be studied in a gauge-invariant way, by 
calculating the effective action $W=\int d^4 x {\cal L}_{\rm eff}$. 
It is represented 
by one-loop diagrams that can be obtained 
by gluing together the external legs of the diagrams 
in Fig. 2. The calculation can be performed by using 
Schwinger's method \cite{schw}, which is often viewed as the 
clearest proof that a constant background electric field 
causes particle production. Since 
$\langle 0\; {\rm out}|0\; {\rm in}\rangle =\exp(iW)$, the quantity 
\begin{equation}\label{prob}
|e^{iW}|^2 =e^{-2{\rm Im}W}
\end{equation} 
represents the probability that no actual pair creation occurs. 
Schwinger's method leads to an exact expression for 
${\cal L}_{\rm eff}$ for the case when the 
background field $\bar{F}_{\mu\nu}$ is constant. For this 
case, his method gives that ${\rm Im}{\cal L}_{\rm eff}=0$ when   
$\bar{{\bf E}}\cdot\bar{{\bf B}}=0$ and 
$\bar{{\bf E}}^2-\bar{{\bf B}}^2<0$, whereas 
${\rm Im}{\cal L}_{\rm eff}>0$ when 
$\bar{{\bf E}}\cdot\bar{{\bf B}}=0$ and 
$\bar{{\bf E}}^2-\bar{{\bf B}}^2>0$. The latter case corresponds to 
an unstable vacuum, i.e. to particle 
creation. However, there are several weaknesses of the method 
presented in \cite{schw}. 
First, the method expresses ${\cal L}_{\rm eff}$ as a 
certain integral over a real variable $s$. This ${\cal L}_{\rm eff}$ 
is divergent and {\em real}. Of course, the divergence is 
expected because it is a one-loop calculation. In this method, 
the divergences are related to the poles of the subintegral 
function. Schwinger obtained the imaginary contributions in an 
artificial way, by replacing the contour of integration with a 
contour that avoids the poles.
Second, a pole exists even for the case 
$\bar{F}_{\mu\nu}=0$. Schwinger removed this 
undesirable term by hand, by introducing an appropriate 
counter-term. If the addition of such counter-terms is allowed, 
then one can also introduce a counter-term that will cancel  
the contribution to ${\rm Im}{\cal L}_{\rm eff}$ that comes 
from finite $\bar{F}_{\mu\nu}$. Such a renormalization seems to be the 
most natural and leads to a stable vacuum. 
Third, if $\bar{F}_{\mu\nu}(x)$ is not constant, such that 
the Fourier transform $\bar{F}_{\mu\nu}(k)$ possesses contributions 
from $k^2>4m^2$, then ${\rm Im}W$ can be {\em negative} 
(at least at the lowest order of perturbation theory).  
Schwinger derived it (see (6.33) in \cite{schw}), but did not 
comment it. This is a pathological result because it leads 
to the result that the probability that no actual pair creation occurs 
can be larger than 1. This pathology suggests again that the 
contour of integration should remain on the real axis or that 
the counter-terms should be introduced for all imaginary 
contributions. It is not consistent to introduce the 
counter-terms only for negative contributions to ${\rm Im}W$.  

At this point, it is instructive to discuss other methods 
that predict particle production by a background source 
$J_{\rm int}$. The most popular method is the Bogoliubov 
transformation. It is important to note that this method 
is applicable only to systems that can be described by  
Lagrangians of the form (\ref{quadr}) and is equivalent 
to the squeezed-state method \cite{gris}.  
In this method, the definition of particles 
is based on the identification of positive-frequency 
solutions to the equation of motion.
The particle production described in this way is also 
inconsistent with energy conservation \cite{nikolic}. 
Another problem with the Bogoliubov-transformation
method is the fact that the 
identification of the positive-frequency solutions to the
classical equations of motion depends on 
the choice of coordinates; different
choices of the time coordinate lead to different particle
contents of the same quantum state \cite{bd}. It is often
argued that this does not make the theory inconsistent, because
different observers have different natural choices of the
time coordinate, so different observers observe different
particle contents. However, in the case of an electromagnetic 
background,
the problem becomes even more serious, because
the identification of the positive-frequency solutions depends
also on the choice of gauge \cite{padmprl}.
In particular, for the case $\bar{F}_{\mu\nu}=0$, the 
``natural" gauge is $\bar{A}_{\mu}=0$, which defines the 
``natural" vacuum. By using a different gauge one can 
obtain that this vacuum is a many-particle state. 
An appropriate gauge leads to a thermal distribution of 
particles in the ``natural" vacuum, leading to an 
electromagnetic analog of the Unruh effect. On the other hand, 
physical quantities should not depend on gauge. All this
suggests that the Bogoliubov transformation is not a 
well-founded method for studying particle creation.    

Finally, let us note that particle production by an 
electromagnetic background can be viewed as a 
tunelling process \cite{paren1,padmprd}. In general, this 
method is also gauge dependent. 

Before drawing a conclusion, let us shortly discuss a more general 
case, in which $J$ in (\ref{quadr}) can be a quantum-field  
operator. Our discussion above does not refer to such a case. 
The most interesting example is the spinor 
quantum electrodynamics, in which the full 
quantum-field operator $\hat{A}_{\mu}$ can be written as 
\begin{equation}\label{full}
\hat{A}_{\mu}=\bar{A}_{\mu}+A_{\mu} \; ,
\end{equation}
where $\bar{A}_{\mu}$ is a c-number background and $A_{\mu}$ 
is an operator with the property $\langle 0|A_{\mu}|0\rangle =0$. 
Such an interaction allows physical processes that are forbidden 
when the electromagnetic background described by $\bar{A}_{\mu}$ 
is not present. For example, the pair production 
$\gamma\rightarrow e^+ e^-$ becomes kinematically allowed 
because the massive object that generates $\bar{A}_{\mu}$ 
can recoil. Such a process has been measured
\cite{wrig} and is not in contradiction with our theoretical 
results that suggest that the process ${\rm vacuum}\rightarrow e^+ e^-$
is forbiden. The latter process has never been measured.        

As we have seen, all methods that predict unstable vacuum in 
systems described by a Lagrangian of the form (\ref{quadr}) 
suffer from certain ambiguities related to the choice of 
coordinates, gauge, or counter-terms. Besides, all methods 
have problems with the conservation of energy. This suggests 
that we should define particles in a general covariant and  
gauge-invariant way and that this definition should automatically 
imply that the number of particles is conserved 
for any system described by a Lagrangian of the form 
(\ref{quadr}), where $J$ is not a quantum-field operator. Since it is 
usual to define the concept of particles by the 
``free" Lagrangian that conserves the number of particles, 
our discussion suggests that any Lagrangian of the form 
(\ref{quadr}) (where $J$ is not a quantum-field operator) 
should be viewed as a ``free" Lagrangian. In particular, this 
will provide the general covariance and gauge 
invariance of the concept of particles because (\ref{quadr}) 
is invariant with respect to general coordinate transformations 
and gauge transformations. If the separation of (\ref{quadr}) 
into the ``free" and ``interacting" part is made in the usual way, 
then the separate parts are not  
general invariant and gauge invariant. Note also that a gauge 
transformation $\hat{A}_{\mu}\rightarrow \hat{A}_{\mu} +
\partial_{\mu}\lambda$, where $\lambda$ is a 
c-number function, can always be written as
\begin{eqnarray}\label{gaugetr}
& \bar{A}_{\mu}\rightarrow \bar{A}_{\mu}+\partial_{\mu}\lambda \; , &
  \nonumber \\
& A_{\mu}\rightarrow A_{\mu} \; ,
\end{eqnarray}
which means that the interaction Lagrangian 
${\cal L}_{\rm int}=-eA_{\mu}\bar{\psi}\gamma^{\mu}\psi$ is 
gauge invariant with respect to such gauge transformations. 
In addition, our discussion suggests that
one has to reject a common belief that the definition   
of particles should be closely related to the definition of positive
frequencies. The explicit definition of particles that obeys 
all properties suggested in this paragraph will be given 
elsewhere.  

\vspace{0.5cm}
\noindent   
{\bf Acknowledgement}
\vspace{0.5cm}

This work was supported by the Ministry of Science and Technology of the
Republic of Croatia under Contract No. 00980102.

\newpage 


\begin{figure}\label{Fig.1}
\caption{The elementary vertex related to the interaction 
of the form ${\cal L}_{\rm int}=\phi^{\dagger}J_{\rm int}\phi$, 
where $J_{\rm int}$ is not a quantum-field operator. }
\end{figure}

\begin{figure}\label{Fig.2}
\caption{The diagram with two external legs calculated at 
all orders of perturbation theory for the interaction as in 
Fig. 1.}
\end{figure}

\newpage
\mbox{}


\begin{figure}[c]
\centerline{\includegraphics{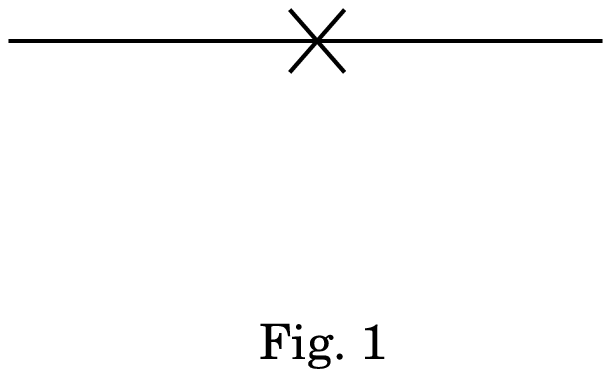}}
\end{figure}

\begin{figure}
\centerline{\includegraphics{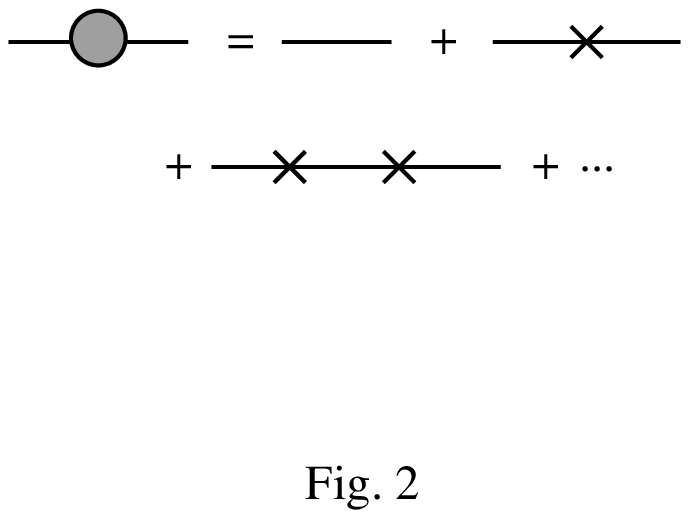}}
\end{figure}


\begin{thebibliography}{9}
\bibitem{schw}
J. Schwinger, Phys. Rev. 82 (1951) 664.
\bibitem{bd}
N. D. Birrell and P. C. W. Davies, Quantum fields in
curved space (Cambridge Press, NY, 1982).
\bibitem{ryder}
L. H. Ryder, Quantum Field Theory (Cambridge University Press, 
Cambridge, 1984).
\bibitem{gris}
L. P. Grishchuk and Y. V. Sidorov, Phys. Rev. D 42 (1990) 3413.
\bibitem{nikolic}
H. Nikoli\'c, hep-th/0103053.
\bibitem{padmprl}
T. Padmanabhan, Phys. Rev. Lett. 64 (1990) 2471.
\bibitem{paren1}
R. Brout, R. Parentani and Ph. Spindel, Nucl. Phys. B 353 (1991) 209.
\bibitem{padmprd}
K. Srinivasan and T. Padmanabhan, Phys. Rev. D 60 (1999) 024007.
\bibitem{wrig}
L. E. Wright, K. K. Sud and D. W. Kosik, Phys. Rev. C 36 (1987) 562.

\end{thebibliography}
\end{document}